# Adverse effects of vaccinations against the Corona-virus SARS-CoV-2: insights and hindsights from a statistical perspective

Thomas Schmidt-Melchiors[1], Jens-Peter Kreiss[2] and Alexander Braumann[3]
Institut für Mathematische Stochastik
TU Braunschweig
Universitätsplatz 2
D-38106 Braunschweig

[1]T.Schmidt-Melchiors@email.de
[2]j.kreiss@tu-braunschweig.de
[3]a.braumann@tu-braunschweig.de

March 10, 2022

**Abstract**

*Vaccinations against the virus SARS-CoV-2 have proven to be most effective against a severe corona disease. However, a significant minority of people is still critical of such a vaccination or even strictly reject it. One but surely not the only reason for this is the fear of undesired adverse effects. During the year 2021 in Germany alone approximately 150 million vaccination doses have been applied. This should be a solid basis to assess the risk of adverse effects of corona vaccinations. Based on publicly available data, especially from the German Federal Institute of Vaccines and Biomedicines (Paul-Ehrlich-Institut) and further scientific publications from Europe, Israel and the United States, this paper tries to give sound quantitative statements on the risk of (severe) adverse effects (e.g. myocarditis, thrombosis and thrombocytopenia, venous thrombosis, including cerebral venous sinus thrombosis and fatalities based thereon) of the various vaccination types, i.e. Comirnaty, Vaxzevria, Janssen and the Spikevax Covid-19 vaccine. The paper also describes some quite serious concerns about the lack of details in the publicly available data and the variations in the structure of reports from the primary source that prohibit even simple time series comparisons.*

*However, from an overarching view all provided vaccines are pretty safe and fall mostly into the same frequency categories with respect to various clusters of undesired complications. Nevertheless there are statistically significant differences amongst the four vaccines which indicate that Vaxzevria is according to some important aspects only the second best preferable choice.*

*The paper aims to support a rational approach regarding the confidence in the widely available SARS-CoV-2 vaccines through a purely statistical investigation of the number of adverse effects in an unbiased manner.*

# 1. Introduction

Vaccinations against the Corona-virus SARS-CoV-2 are the most effective means against the Corona pandemic. Fortunately enough four effective vaccines are available in Germany: The mRNA-based vaccines Comirnaty from BioNTech/Pfizer and Spikevax from Moderna and the vector-based vaccines Vaxzevria from AstraZeneca and Janssen from Johnson&Johnson. Since the end of December 2020 more than 147 Million doses have been applied in total leading to a rate of more than 74% of fully vaccinated persons in the German population. In addition to those parts of the population which cannot be vaccinated due to medical reasons or the non-availability of approved vaccines for children below five years there is a significant percentage of people that are either sceptic against the vaccinations or oppose the vaccinations more as a matter of principle.

Besides the fact that it is unique in history that those vaccines were developed and thoroughly tested in such a short time period it is even more spectacular that a gigantic field test with billions of vaccinations has taken place during the last 12 months. This novelty in itself might create scepticism, but other sources are the numerous reports of adverse effects accompanying the vaccinations campaigns and in particular in Germany some swings in the recommendations of the permanent vaccination committee (Ständige Impfkommission, in short STIKO), an independent scientific and medical consulting body, specifically with respect to the application of Vaxzevria (STIKO 2021-a, STIKO 2021-b, STIKO 2021-c, STIKO 2021-d).

However, there have been numerous articles published on the efficacy, safety and potential adverse effects of the Covid-19 vaccines stemming from different countries.

Before addressing the issues and methodologies examined in this paper, we provide a brief review of relevant literature on the topic of adverse reactions to the licensed COVID-19 vaccines. A list of COVID-19 vaccines licensed in Europe can be found in Meyer (2021). Novak, Tordesillas and Cabanillas, B. (2021) describe the current status on the safety and adverse side effects of COVID-19 vaccines in a very general way.

Detailed results from studies on severe adverse events, especially thrombocytopenia and cerebral venous thrombosis after vaccination with AstraZeneca's vaccine can be found in Greinacher et al. (2021).

For health data essentially stemming from the U.S., Taquet et al. (2021) compare risks of portal vein thrombosis and cerebral venous sinus thrombosis for individuals vaccinated with either of Pfizer-BioNTech's or Moderna's mRNA vaccines on the one hand and patients with proven COVID-19 infections on the other in a very large statistical study. Although the respective risks are small, the study confirms that the risk of both cerebral venous sinus thrombosis and portal vein thrombosis were significantly higher in the COVID-19 group. The sample size is large, with 537,913 individuals affected by COVID-19 and 389,034 individuals vaccinated with a mRNA vaccine. Corresponding comparisons for AstraZeneca's vector-based vaccine could not be made in the study by Taquet et al. (2021) because this vaccine was not used in the United States.

A comparable and also rather large study in the United Kingdom is described in Menni et al. (2021). This study deals with vaccine adverse effects of Pfizer-BioNTech's mRNA vaccine and AstraZeneca's vector-based vaccine. One finding of the study is that vaccine adverse effects occurred less frequently in the available data than would have been expected from the phase III trials. Furthermore, due to the detailed data structure on age, gender, concomitant diseases, health care occupation, etc., the study by Menni et al. (2021) allows for individualized prediction of adverse events based on age, gender and previous COVID-19 status.

Recently Barda et al. (2021) analysed the safety of BioNTech's mRNA vaccine for a huge study of vaccinated and control groups each included nearly one million persons. The underlying detailed data stem from a large health care organization in Israel and they allow for a high resolution with respect to an impressing number of covariables. The obtained results once again underpin the high safety of BioNTech's mRNA vaccine.

A report of myocarditis in adolescents after full vaccination with Comirnaty from Pfizer-BioNTech can be found in Marshall et al. (2021).

The adverse occurrence of thrombosis and thrombocytopenia after vaccination with AstraZeneca is also reported in a paper by Schultz et al. (2021).

Pottegard et al. (2021) described the risk for venous thromboses in Denmark and Norway and Burn et al. (2021) investigated the risk of thromboses for both Vaxzevria and Comirnaty in Spain.

Finally, a detailed presentation of a total of 269 cases of venous thrombosis, including cerebral venous sinus thrombosis, occurring in several European countries after COVID-19 vaccination with AstraZeneca can be found in a March 2021 report by the European Medicines Agency (2021).

In a preceding article Schmidt-Melchiors and Kreiss (2021) have undertaken a statistical analysis of the of various adverse side effects based on German data from the Paul-Ehrlich-Institut and also dwelled on the limitations caused by insufficient details available from the underlying data.

In this article we aim to examine whether the statistical conclusions from our previous article are still valid in the light of extended data available now. We furthermore want to compare our results with those from Austria and Israel as far as these are available. In addition to that we include recent adverse effects like the myocarditis, which has become prominent. Last but not least we will find out whether the quality of data available in Germany has improved during the last months.

The main basis of the analyses in this paper are the summary data on vaccination numbers and adverse events as published in several safety reports of the Paul-Ehrlich-Institut (PEI) (see Paul-Ehrlich-Institut (2021a)-(2021f)). These data are by no means as detailed as the data on which the publications described above by Taquet et al. (2021), Menni et al. (2021) and Barda et al. (2021) are based. This is by and large also true for the available Austrian data provided by the Bundesamt für Sicherheit im Gesundheitswesen – BASG (2021).

# 2. Initially analysed topics, basic data from Germany, Austria and Israel and methodology in general

In Schmidt-Melchiors and Kreiss (2021) we tried to answer the following question based on statistical analyses:

1. Are the frequencies of occurrence of adverse effects overall significantly different for the applied vaccines?

2. Are the frequencies of occurrence of those severe adverse effects which cause permanent damages or fatality significantly different for the applied vaccines?

3. Are the frequencies of occurrence of the severe adverse effects of venous thromboses or embolism significantly different for the applied vaccines?

4. Are the frequencies of occurrence of the severe adverse effects of specific brain vein thromboses with and without acute thrombocytopenia significantly different for the applied vaccines?

For this article we will add the question

5. Are the frequencies of occurrence of the severe adverse effects of myocarditis significantly different for the applied vaccines?

Whenever the data situation permits plausible further detailing with respect to gender or age we will incorporate this into the analysis.

Last but not least we will investigate whether the described results are in line with findings from other countries, i.e. Austria and Israel.

The overall situation in Germany is described on the basis of the so called security reports from the Paul-Ehrlich-Institut (www.pei.de), namely those from 9. 4. 2021, 7. 5. 2021, 10. 6. 2021, 19.8.2021 and 20.9.2021 (cf. Paul-Ehrlich-Institut (2021-a)-(2021-f)).

| Number of applied vaccination doses | 2 April 2021 | | | 29 April 2021 | | | 31 May 2021 | | | 1 August 2021 | | | 31 August 2021 | | | 30 September 2021 | | | |
|---|---|---|---|---|---|---|---|---|---|---|---|---|---|---|---|---|---|---|---|
| vaccination product | Total | 1. Dose | 2. Dose | Total | 1. Dose | 2. Dose | Total | 1. Dose | 2. Dose | Total | 1. Dose | 2. Dose | Total | 1. Dose | 2. Dose | Total | 1. Dose | 2. Dose | Booster |
| Reporting period (data lock point) | 02/04/21 | | | 29/04/21 | | | 31/05/21 | | | 01/08/21 | | | 31/08/21 | | | 30/09/21 | | | |
| Comirnaty | 10,722,876 | 6,562,591 | 4,160,285 | 21,329,667 | 15,394,941 | 5,934,726 | 36,865,276 | 24,120,063 | 12,745,213 | 68,962,481 | 35,492,080 | 33,470,401 | 76,982,568 | 37,831,661 | 39,150,907 | 82,341,579 | 39,739,132 | 41,864,164 | 738,283 |
| Spikevax | 713,067 | 541,246 | 171,821 | 1,667,261 | 1,266,702 | 400,559 | 3,972,764 | 2,880,517 | 1,092,247 | 8,506,260 | 4,239,829 | 4,266,431 | 9,396,381 | 4,398,792 | 4,997,589 | 9,668,138 | 4,460,412 | 5,177,336 | 30,390 |
| Vaxzevria (AZ) | 2,945,125 | 2,943,061 | 2,064 | 5,775,546 | 5,731,540 | 44,006 | 9,230,103 | 8,530,534 | 699,569 | 12,491,937 | 9,190,193 | 3,301,744 | 12,645,915 | 9,214,941 | 3,430,974 | 12,692,700 | 9,237,104 | 3,455,596 | 0 |
| Covid19-Vaccine Janssen | | | | 2,106 | N/A | 2,106 | 472,941 | N/A | 472,941 | 2,416,109 | N/A | 2,416,109 | 2,852,260 | N/A | 2,852,260 | 3,186,297 | N/A | 3,185,598 | 699 |
| Totals | 14,381,068 | 10,046,898 | 4,334,170 | 28,774,580 | 22,393,183 | 6,379,291 | 50,541,084 | 35,531,114 | 14,537,029 | 92,376,787 | 48,922,102 | 41,038,576 | 101,877,124 | 51,445,394 | 47,579,470 | 107,888,714 | 53,436,648 | 50,497,096 | 768,673 |

Table 1: Number of applied vaccinations per vaccine at various points in time (data lock points)
*Sources are from „Tabelle 1" of the according „PEI Sicherheitsbericht" from 9.4., 7.5., 10.6., 19.8., 20.9. and 26.10.2021 respectively*

We can see that until end of August 2021 in total more than 101 million doses have been administered. In the following month only about six million additional doses have been applied. The relative portion become more visible in the following graphical presentation:

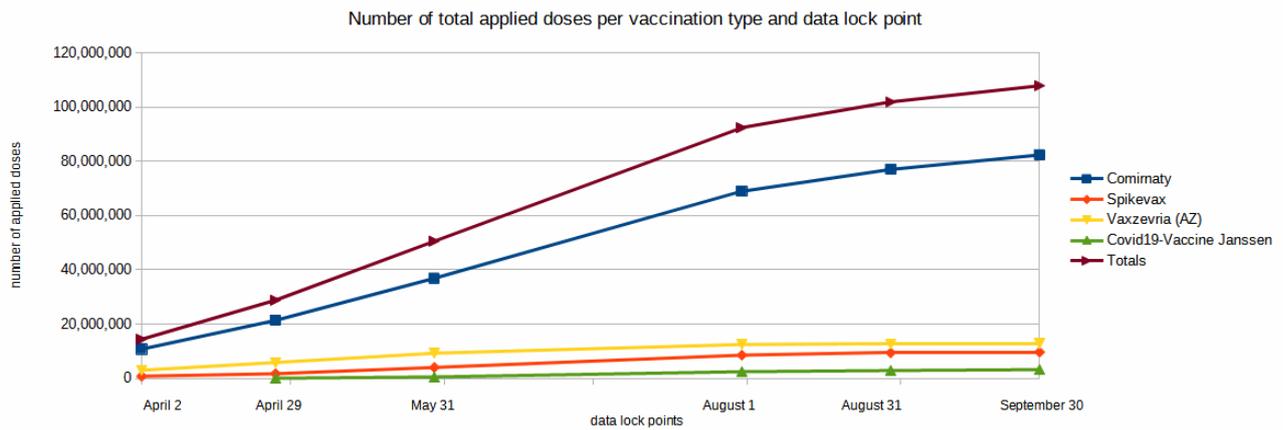

Diagram 1: Progress of number of vaccinations per vaccine over time

Obviously by far the dominating applied vaccine is Comirnaty from BioNTech/Pfizer with roughly 80% of the total of applied doses. Vaxzevria from AstraZeneca is still the second most often administered vaccine but Spikevax from Moderna is closing up, in particular as quite a number of originally planned second doses of Vaxzevria have been replaced by Comirnaty and Spikevax, following a recommendation from the German permanent vaccination committee (STIKO 2021-d). In fact the number of Vaxzevria vaccination will not increase significantly any more as the German government has decided to procure no further lots of Vaxzevria for Germany but instead donate them to the COVAX-Program (Federal Ministery of Health 2021). This coincides with our former conclusion (Schmidt-Melchiors and Kreiss (2021)) that although usually still being in the same safety category with respect to adverse effects of the vaccination Vaxzevria is nevertheless inferior to the alternatives.

The Covid-19 vaccine Janssen is also not far away as one has to bear in mind that only one dose needs to be applied. However, recently the STIKO has recommended that the one application of Janssen should be followed by a booster vaccination with a mRNA-vaccine as early as three weeks after the Janssen vaccination (STIKO 2021-e). This will trigger the question whether due to a lower efficacy Janssen will as well not be procured any more or only in small quantities for specific target groups that need some protection in the short term.

It would be extremely valuable to have a break-down of these numbers per vaccine and data lock point for at least gender and age groups. However, this is still not available, except a broad estimation by the Robert-Koch-Institut (RKI: www.rki.de) for the distribution of gender overall:

| Overall distribution of vaccinations according to gender (August 2021) | percentage |
|---|---|
| male | 48.00 % |
| female | 52.00 % |

Table 2: Overall distribution of vaccinations according to gender up to August 2021
source: Paul-Ehrlich-Institut (2021-e)

Unfortunately no differentiation by type of vaccine is provided. The described ratio shows a slightly higher portion of females than in the total population of Germany which is 50.7% (Statis 2021).

For some adverse effects the Paul-Ehrlich-Institut reports age-specific data. But since there is no correspondent data provided for the underlying vaccinated groups it does not allow to determine age-specific risks.

For Austria our analysis is based on data published by the Austrian Ministry of Social Affairs, Health, Care and Consumer Protection and the Austrian Federal Office for Safety in Health Care, see BMGSPK (2021) and BASG (2021). As far as the calculation of risks is concerned the situation is quite the opposite compared to Germany. While the distribution of vaccinations according to gender, age and vaccine is available (BMSGPK 2021), the numbers on adverse effects are only reported by either vaccine or age. In addition, reported age-groups in BMGSPK (2021) and BASG (2021) are different. Therefore age- and gender-specific risks cannot be determined.

The data from Israel is taken from Barda et al. (2021). Whilst many characteristics are used to get an extremely good match to the two comparison groups (unvaccinated persons / Covid-19-infected persons) these items are not evaluated in a factor analysis.

The probably best statistical approximations for the analysed frequencies are based on the Binomial distribution. The high number of trials would usually justify an approximation by the asymptotic Normal or Poisson distribution with the latter being more intuitive. Schmidt-Melchiors and Kreiss (2021) have shown that in the analysed examples the Poisson distribution assumption resulted in confidence intervals for the expected frequencies that were very close to those provided by a Clopper-Pearson based confidence interval (cf. Hartung, Elpelt and Klösener (2009), Section 3.1.1. in Chapter IV or the original article from Clopper and Pearson (1934)). Due to its robustness we will use the Clopper-Pearson 95%-confidence intervals to determine whether the compared expected values differ significantly from each other.

## 3. Occurrence rates for reported adverse effects

The Paul-Ehrlich-Institut (PEI) collects all reports of undesired adverse effects and also maintains a sub-category for severe adverse effects.

In their security report from October 26, 2021 the following figures are provided:

| Number of reported adverse events from 27.12.2020 until 30.9.2021 | Number of applied vaccination doses | Total reported events | Respective reporting rate per 1000 vaccinations | Number of reported severe adverse events | Respective reporting rate per 1000 vaccinations |
|---|---|---|---|---|---|
| Comirnaty | 82,341,579 | 94,281 | 1.14 | 12,939 | 0.157 |
| Spikevax | 9,668,138 | 25,713 | 2.66 | 1,493 | 0.154 |
| Vaxzevria (AZ) | 12,692,700 | 45,718 | 3.60 | 5,751 | 0.453 |
| Covid19-Vaccine Janssen | 3,186,297 | 6,243 | 1.96 | 560 | 0.176 |
| Unknown | | 773 | | 311 | |
| Totals | 107,888,714 | 172,728 | 1.60 | 21,054 | 0.195 |

*Table 3: Number of reported adverse effects and severe adverse effects from 27.12.2020 until 30.9.2021*

The reporting rates for the four vaccine types appear very different and indeed vary statistically significant as can be seen from the following table:

| 95%-confidence for reporting rates of adverse effects per vaccination product | lower bound | upper bound |
|---|---|---|
| Comirnaty | 1.138 | 1.152 |
| Spikevax | 2.627 | 2.692 |
| Vaxzevria (AZ) | 3.569 | 3.635 |
| Covid19-Vaccine Janssen | 1.911 | 2.008 |

*Table 4: 95%-confidence intervals for the reporting rate for adverse effects per 1000 vaccinations for all four vaccines based on data from 27.12.2020 until 30.9.2021*

The four intervals mutually do not overlap and provide a clear ranking with Comirnaty coming in first, then Janssen and Spikevax and at the end Vaxzevria with the highest rates.

The same result was found in our previous analysis, although the rates itself were slightly higher for Vaxzevria and slightly lower for the other vaccines.

Similar results are also available for Austria (cf. Table A1 in the Appendix).

Based on the gender distribution ratios from Table 2 and assuming a homogeneous distribution across all vaccines we can derive the following table to find out whether there is a dependency of gender on the reporting rates (unfortunately the split according to gender per vaccine is not provided in the latest report).

| 95%-confidence for reporting rates of adverse effects per vaccination product and gender | lower bound males | upper bound males | lower bound females | upper bound females |
|---|---|---|---|---|
| Comirnaty | 0.631 | 0.647 | 1.569 | 1.593 |
| Spikevax | 1.330 | 1.397 | 3.620 | 3.725 |
| Vaxzevria (AZ) | 2.232 | 2.308 | 4.920 | 5.027 |
| Covid19-Vaccine Janssen | 1.834 | 1.973 | 1.893 | 2.028 |

*Table 5: 95%-confidence intervals for the reporting rate for adverse effects per 1,000 vaccinations for all 4 vaccines differentiated by gender for vaccinations until data lock point September 30, 2021*

As can be seen from Austrian data, the assumption of a homogeneous distribution across all vaccines might be not fulfilled (cf. Table A2 in the Appendix).

For all vaccines except Janssen the reporting rates of females are significantly higher. The overall ranking is almost preserved with the impressive exception for Spikevax where the rate for males is the second lowest and thus lower than that for Janssen whilst the reporting rate for females is second highest and the spread between male and female is almost the factor three. Overall there are possible explanations like a higher responsive immune system for women, in particular as the applied quantity is the same for men and women but the per kg-weight-ratio is thus in average higher than for men. It could also be that women are more open to report side effects. However, part of the explanation can also still lie in a deviation from the assumption of evenly spread female-to-male ratios across the vaccines.

More important than the reporting rate based on the total sum of reported adverse effects should be the analogous rate for severe adverse effects. Here the situation is as follows:

| 95%-confidence for reporting rates of severe adverse effects per vaccination product | lower bound | upper bound |
|---|---|---|
| Comirnaty | 0.154 | 0.160 |
| Spikevax | 0.147 | 0.162 |
| Vaxzevria (AZ) | 0.441 | 0.465 |
| Covid19-Vaccine Janssen | 0.161 | 0.191 |

*Table 6: 95%-confidence intervals for the reporting rate for severe adverse effects per 1,000 vaccinations for all four vaccines based on data from 27.12.2020 until 30.9.2021*

These reporting rates are roughly at least ten times lower than the reporting rates for all reported adverse effects and they are statistically hardly significantly distinguishable except for Vaxzevria which has about three times higher rates than the other vaccines. This is pretty much in line with our previous results except that Janssen that showed the lowest rates until end of May has not only closed up to the two mRNA-vaccines but its lower 95% confidence bound is almost exceeding the upper bounds of the mRNA-vaccines. This does not come totally unexpected as until end of May Janssen had been applied for a much shorter period and in a markable lower number of vaccinations.

It should be noted however that all these reporting rates are prone to an underestimation as there is usually a latency between reporting of application of vaccine and reporting of adverse effects which might span even up to ca. 40 days. It would be worthwhile to calculate adjusted estimations to account for this latency and monitor them on a timely bases to be able to detect any trends. However we believe that in the current situation this bias is neither changing the relative order nor the order of magnitude. We will dwell on this further when dealing with the most important categories of fatal effects or persisting damages.

# 4. Occurrence of extreme severe adverse effects

A major concern for many people are extreme severe adverse effects like persisting health damages or even fatalities that occurred in a short time span after the vaccination.

These are categories which are explicitly mentioned in the security report from the Paul-Ehrlich-Institut from 26.10.2021 (cf. Paul-Ehrlich-Institut (2021-f)). However, since the case numbers can only be calculated from stated fractions of the overall incidents per vaccine and these fractions are rounded it happened for the Comirnaty vaccine that the calculated numbers became less between two data lock points. Although theoretically the effect could stem from corrected data we are inclined to believe that these are simple rounding effects and have chosen adjust the numbers in this case for the subsequent lock point to the same value as for the foregoing one. The rounding effect may also be the cause that the total number of reported fatalities (1,809) is higher than the sum of calculated numbers for the four different vaccines (1,718) unless the difference is not completely caused by those events which could not be assigned to a specific vaccine (category "unknown"). Keeping all this in mind the following numbers were derived:

| Number of reported deaths and persisting damages and their respective reporting rates per 100000 applied vaccinations from 27.12.2020 until 30.9.2021 | Number of applied vaccines | cumulated number of deaths | corresponding reporting rate | cumulated number of persisting damages deaths | corresponding reporting rate | cumulated number of both reported events types | corresponding reporting rate |
|---|---|---|---|---|---|---|---|
| Comirnaty | 82,341,579 | 1,271 | 1.54 | 2,639 | 3.20 | 3,910 | 4.75 |
| Spikevax | 9,668,138 | 51 | 0.53 | 462 | 4.78 | 513 | 5.31 |
| Vaxzevria (AZ) | 12,692,700 | 365 | 2.88 | 914 | 7.20 | 1,279 | 10.08 |
| Covid19-Vaccine Janssen | 3,186,297 | 31 | 0.97 | 143 | 4.49 | 174 | 5.46 |

*Table 7: Number of reported deaths and persisting damages and their respective reporting rates per 100,000 applied vaccinations from 27.12.2020 until 30.9.2021*

These numbers lead to the following 95%-confidence intervals for the three respective reporting rates:

| 95%-confidence levels for reporting rates of cases of persisting damages or fatal events per 100000 vaccinations and per vaccine | lower bound cases of persisting damages | upper bound cases of remaining damages | lower bound fatal events | upper bound fatal events | lower bound both events | upper bound both events |
|---|---|---|---|---|---|---|
| Comirnaty | 3.084 | 3.330 | 1.460 | 1.631 | 4.601 | 4.900 |
| Spikevax | 4.353 | 5.235 | 0.393 | 0.694 | 4.857 | 5.786 |
| Vaxzevria (AZ) | 6.742 | 7.683 | 2.588 | 3.186 | 9.532 | 10.644 |
| Covid19-Vaccine Janssen | 3.783 | 5.287 | 0.661 | 1.381 | 4.680 | 6.335 |

*Table 8: 95%-confidence intervals for the reporting rate for extreme severe adverse effect categories per 100,000 vaccinations based on data from 27.12.2020 until 30.9.2021*

Considering the combined reported rates we see the already observed clear dichotomy with Vaxzevria showing significantly higher results than the other three vaccines for which the rates are statistically not to be differentiated at the 95%-confidence level.

Results for Austria do not show such a significant difference in the rates of reported deaths but the 95%-confidence intervals are in general much wider (cf. Table A3 in the Appendix).

Looking the cases of persisting damages in isolation the situation is slightly different as Comirnaty provides statistically the lowest results although not far away from Janssen which in turn is pretty close to Spikevax. This situation is basically reversed for these three vaccine types with regards to the category of fatal events: Spikevax is now providing the lowest results although statistically not significantly different from Janssen while Comirnaty comes next and then Vaxzevria with the highest rates. Unfortunately there is no age distribution available for the application of vaccines per type which would allow to compare these rates with the expected death rates in a given adequate time period of e.g. 40 days.

A comparison of these results to those analysed in a previous paper of Schmidt-Melchiors and Kreiss (2021) is illustrated in diagram 2. The situation has not changed significantly for Comirnaty and Spikevax, i.e. the reporting rates are alike or slightly lower for Spikevax with confidence intervals becoming narrower as to be expected. The situation for Vaxzevria has however dramatically changed insofar as the reporting rates increased by a factor three compared to end of April. Although this still means the rate is in the same order of magnitude compared to the others but in a relative comparison it supports the decision of the STIKO and government to prefer the other vaccine types from August onwards.

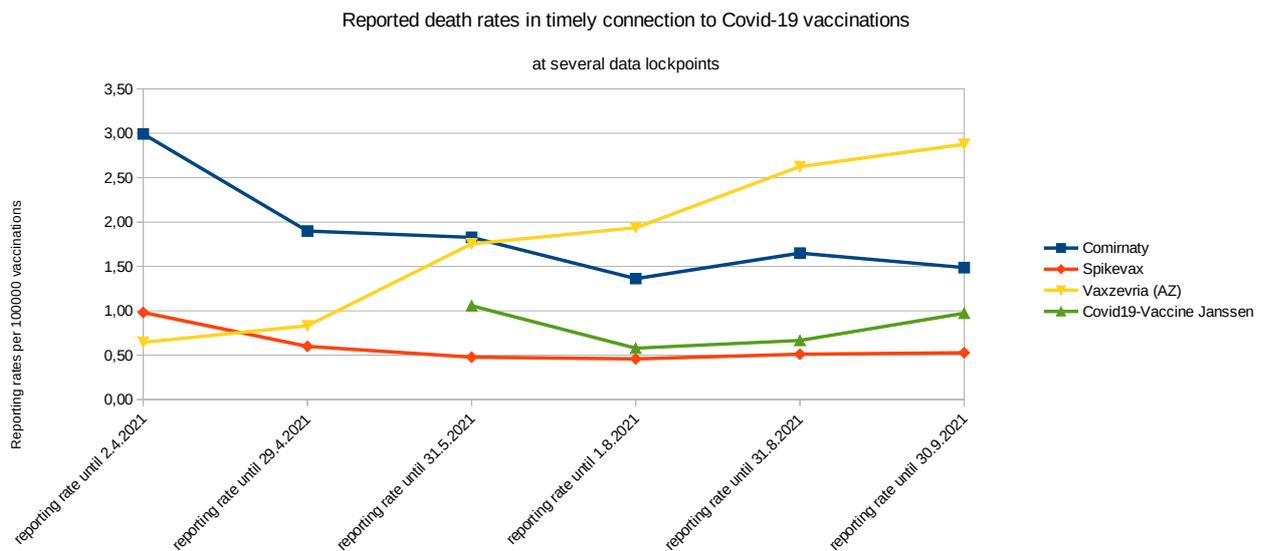

*Diagram 2: Development of reporting rates for fatal events from April 2 to August 31, 2021*

A standard critique to the way adverse effects are reported is that the reporting rates are calculated by the ratio of number of adverse effects divided by the number of vaccinations until the same data lock point. As the adverse effects usually occur with a delay, in particular for the serious ones like persisting damages or fatalities, this results in a systematic underrating. The longer the vaccination campaign stays on, the lesser this underestimation will become. We can demonstrate this by comparing adjusted rates which are derived by dividing the number of adverse effects by the number of vaccinations 20-30 days before, i.e. introducing an artificial time gap allowing for a potential latency between vaccination date and reporting date of the adverse effect. Whilst for the time until end of May these adjusted rate where about two times higher than the original ones, cf. Schmidt-Melchiors and Kreiss (2021), we now find adjusted rates that are just up to ca. 10% higher than the original ones as can be seen from the table below:

| Number of reported deaths until 30.9.2021 and reporting rates per 100000 based on number of applied vaccinations until 30.9.2021 versus until 31.8.2021 | cumulated number of deaths until 30.9.2021 | reporting rate according to PEI until 30.9.2021 | cumulated number of deaths until 30.9.2021 | adjusted reporting rate based on number of vaccinations only until 31.8.2021 |
|---|---|---|---|---|
| Comirnaty | 1,271 | 1.49 | 1,271 | 1.65 |
| Spikevax | 51 | 0.53 | 51 | 0.54 |
| Vaxzevria (AZ) | 365 | 2.88 | 365 | 2.89 |
| Covid19-Vaccine Janssen | 31 | 0.97 | 31 | 1.09 |

*Table 9: original and adjusted reporting rates for most recent fatality numbers based on an assumed 30 days latency period*

These underestimations of in average less than 10% are probably partly compensated by an underestimation of the number of vaccinations as recently indicated by the Robert-Koch-Institut (RKI). Insofar we have now reached a situation where this bias seems nearly negligible.

## 5. Occurrence of thromboses and embolisms

Thromboses and embolisms are severe adverse effects that might end in fatalities, although these are not the only adverse effects that might have fatal consequences.

The occurrence of specific venous thromboses in particular for persons who received a Vaxzevria dose has obtained high attention in the media. One of the earliest scientific publications has been provided by Greinacher et al. (2021), who analysed in detail cases of cerebral venous thromboses in conjunction with a thrombocytopenia. These and other similar analysis eventually led to a significant change in the guidelines of the STIKO that advised against a Vaxzevria vaccination for persons under the age of 60 years.

In general the situation is differently described in various studies from abroad. In Denmark and Norway Vaxzevria was sorted out after e.g. Pottegard et al. (2021) described an increased risk for venous thromboses. This is in line with results from Schmidt-Melchiors and Kreiss (2021), where Vaxzevria was found to have a significant higher risk than the other vaccines. On the other hand Burn et al. (2021) based on data from Catalonia (Spain) looking at Vaxzevria and Comirnaty found statistically similar increased risk profiles for both vaccines for venous thromboses (VTE) but not for arterial thromboses (ATE) and thromboses with thrombocytopenia (TTS) but remarked that the relative risk for VTE after an infection with Covid-19 was about six times higher than after the first vaccination with Comirnaty. Somewhat different results are reported by Simpson et al. (2021) based on data from Scotland. Whilst a first dose application of Comirnaty was not associated with any statistically significantly increased risk for VTE, ATE, TTS or ITP (idiopathic thrombocytopenic purpura) the application of Vaxzevria was associated with a small increased risk for ATE, ITP and haemorrhagic events but not with VTE. Similarly Barda et al. (2021) found no increased thrombotic risk for the application of Comirnaty as detailed later.

The security reports from the Paul-Ehrlich-Institut only provide somewhat detailed data until the data lock point August 1$^{st}$, 2021, and even these without discriminating venous and arterial thromboses (Sicherheitsbericht des Paul-Ehrlich-Instituts (2021-d)). They differentiate however between thromboses with and without thrombocytopenia. We therefore consider these categories individually and combined (cf. Table A4 in the appendix). In all cases we also looked for those subsets that number the fatalities in each category. It has to be kept in mind however that there is quite some reporting uncertainty involved, in particular with respect to the definition of a manifest thrombocytopenia. To be on the conservative side we have included the very limited number of mentioned cases even when in doubt about the adequate classification.

This lead to the following results for the 95%-confidence intervals for thromboses without thrombocytopenia:

| 95%-confidence limits for reporting rates for thromboses without thrombocytopenia and fatal events thereof (data until 1.8.2021) | lower limit all thromboses without thrombocytopenia per 100000 vaccinations | upper limit all thromboses without thrombocytopenia per 100000 vaccinations | lower limit thromboses without thrombocytopenia but with fatal ending per 100000 vaccinations | upper limit thromboses without thrombocytopenia but with fatal ending per 100000 vaccinations |
|---|---|---|---|---|
| Comirnaty | 2.431 | 2.671 | 0.226 | 0.304 |
| Spikevax | 1.873 | 2.512 | 0.056 | 0.216 |
| Vaxzevria (AZ) | 9.568 | 10.692 | 0.409 | 0.672 |
| Janssen | 2.222 | 3.614 | 0.045 | 0.424 |

Table 10: 95%-confidence intervals for the reporting rates for thromboses without thrombocytopenia per 100,000 vaccinations based on data from 27.12.2020 until 1.8.2021

It is obvious that there is a clear spike for the reporting rates of Vaxzevria: for the thromboses overall it is a factor 4-5 higher than the others and for the fatal thromboses it is still a factor 2-3 higher. The other three vaccines play in the same league although Janssen is slightly higher which should be monitored as the number of vaccinations is still pretty low compared to the other vaccines.

We will now look at the situation where thromboses occurred together with a thrombocytopenia and show the corresponding results in the next table.

| 95%-confidence limits for reporting rates for thromboses with thrombocytopenia and fatal events thereof (data until 1.8.2021) | lower limit all thromboses with thrombocytopenia per 100000 vaccinations | upper limit all thromboses with thrombocytopenia per 100000 vaccinations | lower limit thromboses with thrombocytopenia but with fatal ending per 100000 vaccinations | upper limit thromboses with thrombocytopenia but with fatal ending per 100000 vaccinations |
|---|---|---|---|---|
| Comirnaty | 0.005 | 0.023 | 0.000 | 0.008 |
| Spikevax | 0.003 | 0.085 | 0.000 | 0.043 |
| Vaxzevria (AZ) | 1.194 | 1.616 | 0.142 | 0.314 |
| Janssen | 0.257 | 0.868 | 0.045 | 0.424 |

Table 11: 95%-confidence intervals for the reporting rates for thromboses with thrombocytopenia per 100,000 vaccinations based on data from 27.12.2020 until 1.8.2021

Again we can see a clear spike for the reporting rates of Vaxzevria: for the thromboses with thrombocytopenia it is about a factor 3 higher than for Janssen and almost two orders of magnitude higher than for the mRNA-vaccines. The situation for those cases with fatal consequences is similar: Vaxzevria and Janssen playing roughly in the same league while the mRNA-vaccine are about two orders of magnitude lower.

It should be noted however that the category of thrombosis with thrombocytopenia contributes for the vector-based vaccines just about 10% and for the mRNA-vaccines just about 1% to the category of all thrombosis. For the latter this is true also for the subcategories of fatal events whilst for the vector-based vaccines the percentage is significantly higher (30-50%). Nevertheless, looking at these numbers the 95%-confidence levels for the combined categories of thromboses the media hype about the sinus venous thrombosis with thrombocytopenia seems somewhat disproportionate. Even for Vaxzevria the odds to die in conjunction with a "normal" thrombosis are about 2.5 times higher than to die in the course of a thrombosis with thrombocytopenia.

| 95%-confidence limits for reporting rates for all thromboses and fatal events thereof (data until 1.8.2021) | lower limit all thromboses per 100000 vaccinations | upper limit all thromboses per 100000 vaccinations | lower limit all thromboses with fatal ending per 100000 vaccinations | upper limit all thromboses with fatal ending per 100000 vaccinations |
|---|---|---|---|---|
| Comirnaty | 2.443 | 2.683 | 0.227 | 0.305 |
| Spikevax | 1.895 | 2.537 | 0.056 | 0.216 |
| Vaxzevria (AZ) | 10.924 | 12.122 | 0.601 | 0.912 |
| Janssen | 2.662 | 4.167 | 0.143 | 0.652 |

*Table 12: 95%-confidence intervals for the reporting rates for all thromboses per 100,000 vaccinations based on data from 27.12.2020 until 1.8.2021*

Summarizing we can conclude that the risk of acquiring a thromboses is overall pretty low, but the relative risk with Vaxzevria is about 4-5 times higher than with the mRNA-vaccines and for the risk for fatal consequences is about three times higher. It would be very helpful to know whether the previous occurrence of thrombosis poses an overall risk for this adverse effects. Unfortunately the data situation available in Germany does not permit such an analysis.

We will not let it unmentioned that there are thrombocytopenia without accompanying thrombosis (idiopathic thrombocytopenic purpura – ITP) reported as adverse effects. The figures for their occurrence and fatality rate are very similar to those for thrombosis with thrombocytopenia in conjunction (cf. Table A4 in the appendix).

Even if we would at this risk to the category of thromboses it can be seen that this can explain only about 30% of all fatal events for e.g. Vaxzevria. This is an important contribution but clearly not the single overwhelming factor.

The above outlined results are by and large supported by the findings in Barda et al. (2021) who investigated the situation in Israel and compared the occurrence of various thrombosis types after Comirnaty vaccination to those in a matching control group of unvaccinated persons.

Their results are shown below:

| Number of reported other thrombosis and thrombocytopenia and respective reporting rates from Israel (Barda et al. -2021) | Total number persons in vaccinated / control group | Number of thrombosis/ thrombocytopenia in vaccination group | Respective reporting rate per 100000 persons | Number of thrombosis/ thrombocytopenia in control group | Respective reporting rate per 100000 persons |
|---|---|---|---|---|---|
| Deep-vein thrombosis | 925,380 | 39 | 4.21 | 47 | 5.31 |
| other thrombosis | 932,469 | 12 | 1.29 | 22 | 2.49 |
| thrombocytopenia | 923,123 | 56 | 6.07 | 60 | 6.78 |
| Sum of all | 923,123 | 107 | 11.59 | 129 | 14.58 |

Table 13: Number of reported thromboses and thrombocytopenia and respective reporting rates per 100,000 persons from Israel (source: Barda et al. (2021))

In principle these rates are somewhat higher even if halved by two to adjust for the fact that the basis is vaccinated persons not vaccinations and assuming in general two vaccinations per persons being applied. On the other hand the numbers show that there is no increase in occurrence compared to the control group indicating that the Comirnaty vaccination is no hazard with respect to these adverse effects. Differences in reporting rates between Israel and Germany might (in part) stem from a significantly different age distribution, however the extremely higher reporting rates for thrombocytopenia in Israel should be further investigated.

# 6. Occurrence of myocarditis and pericarditis

More recently these serious adverse effects have been reported, in particular in the context of mRNA-vaccines (cf. e.g. Marshall et al. (2021)).

Most recent data derived from Sicherheitsbericht des Paul-Ehrlich-Instituts (2021-f) (cf. table below) as well as data from Austria (cf. Tables A6 in the Appendix) show that not only mRNA-vaccines are associated with these adverse effects:

| Number of reported myo-/pericarditis and reporting rates (data 27.12.2020 until 30.9.2021) | Total number of applied doses | Number of myo-/pericarditis | Respective reporting rate per 100000 vaccinations | Number of myo-/pericarditis of males | Respective reporting rate per 100000 vaccinations | Number of myo-/pericarditis of females | Respective reporting rate per 100000 vaccinations |
|---|---|---|---|---|---|---|---|
| Comirnaty | 82,341,579 | 930 | 1.13 | 641 | 1.62 | 289 | 0.67 |
| Spikevax | 9,668,138 | 238 | 2.46 | 184 | 3.96 | 54 | 1.07 |
| Vaxzevria (AZ) | 12,692,700 | 61 | 0.48 | | | | |
| Janssen | 3,186,297 | 27 | 0.85 | | | | |
| Totals | 107,888,714 | 1,256 | 1.16 | | | | |

Table 14: Number of myo-/pericarditis events per vaccine and respective reporting rates per 100,000 vaccinations based on data from 27. 12. 2020 until 30.9. 2021[1] [2]

---

1  The reporting rates for males and females are slightly different than those mentioned in the Tabelle 5 of Sicherheitsbericht des Paul-Ehrlich-Instituts (2021-f) which is probably due to the fact that the male-female-ratio used is slightly different from the overall ratio of 48:52. However, the used ratios have not been disclosed.
2  For Comirnaty there have been 17 cases reported with unknown gender.

For the German data it can be shown however, that the reporting rates for Comirnaty and Spikevax are significantly higher than for Vaxzevria. This cannot be shown so far for Janssen, probably due to the smaller sample size (cf. Table A5 in the Appendix). These findings cannot be supported by data from Austria (cf. Table A6 in the appendix), where the 95%-confidence interval overlap but are also wider.

For the mRNA vaccines a differentiation by gender is available. Using the overall ratio of applied vaccinations by gender from table 2 it can be shown that the occurrence rates for males are significantly higher for both Comirnaty and Spikevax (cf. Table A7 in the Appendix).

The security report from the Paul-Ehrlich-Institut (2021-f) also describes the occurrence frequencies for different age groups (12-17, 18-29, 30-39, 40-49, 50-59, 60-69, 70-79, 80+) and lists reporting rates per age groups and gender[3], although without giving an exact reference for the according stratification of the applied vaccines. Based on that the by far highest reporting rates for both vaccines can be found in age groups between 12 and 29. The preponderance of the effect in the male group is most pronounced in these age groups as well, cf. table below:

| Reporting rate per 100000 vaccinations for myo-/pericarditis per age group and gender | Comirnaty | | Spikevax | |
|---|---|---|---|---|
| | males | female | males | female |
| 12-17 | 4.81 | 0.49 | 11.41 | - |
| 18-29 | 4.68 | 0.97 | 11.71 | 2.95 |
| 30-39 | 1.88 | 1.11 | 4.67 | 1.12 |
| 40-49 | 1.12 | 0.93 | 2.13 | 0.8 |
| 50-59 | 0.71 | 0.77 | 0.99 | 0.91 |
| 60-69 | 0.38 | 0.29 | 0.31 | - |
| 70-79 | 0.47 | 0.25 | 0.5 | 0.45 |
| > 80 | 0.18 | 0.13 | 0.47 | - |
| Total | 1.57 | 0.65 | 3.78 | 1.09 |

*Table 15: Reporting rates for myo-/pericarditis events per 100,000 vaccinations of Comirnaty or Spikevax resp. for different age groups based on data from 27.12.2020 until 30.9.2021 (source table 5 in Paul-Ehrlich-Institut (2021-f))*

The above mentioned report from the Paul-Ehrlich-Institut the reporting rates are significantly higher than those to be expected for a corresponding unvaccinated age group of males for both vaccines and the groups 12-17 and 18-29. However the relative risks for the groups are about 2.5 times higher for Spikevax than for Comirnaty as can be seen in the next table:

---

3   cf. Tabelle 5 of Sicherheitsbericht des Paul-Ehrlich-Instituts (2021-f)

| Relative risk ratio for myo-/pericarditis per age group and gender between Spikevax and Comirnaty | Spikevax vs. Comirnaty | |
|---|---|---|
| | males | female |
| 12-17 | 2.4 | 0.0 |
| 18-29 | 2.5 | 3.0 |
| 30-39 | 2.5 | 1.0 |
| 40-49 | 1.9 | 0.9 |
| 50-59 | 1.4 | 1.2 |
| 60-69 | 0.8 | 0.0 |
| 70-79 | 1.1 | 1.8 |
| > 80 | 2.6 | 0.0 |
| Total | 2.4 | 1.7 |

Table 16: Relative risk ratios based on reporting rates for myo-/pericarditis events per 100,000 vaccinations of Comirnaty or Spikevax resp. for different age groups based on data from 27.12.2020 until 30.9.2021 (source Table 5 in Paul-Ehrlich-Institut (2021-f))

It should be noted that relative risk ratios and absolute reporting ratios need to be judged in conjunction. Hence, the conclusion of the STIKO to not recommend the Spikevax application to males younger than thirty years seems plausible because Comirnaty is the better alternative with respect to this adverse effect although it still provides a signal as well. Unfortunately the report does not mention anything on expected frequencies for the age groups 30-39 years. Since the relative risk was still 2.5 times higher for Spikevax the question whether Comirnaty should be preferred for this age group as well would deserve a dedicated consideration. Although the situation seems similar in the age group over 80 years it is indeed different as there was only just one case in the Spikevax group.

The findings for Germany were again by and large supported by the results of a study from Israel (cf. Barda et al. (2021)) as can be seen by some of those results:

| Number of reported myo-/pericarditis and reporting rates from Israel (cf. Barda et al. - 2021) | Total number persons in vaccinated / control group | Number of myo-/pericarditis vaccination group | Respective reporting rate per 100000 persons | Number of myo-/pericarditis control group | Respective reporting rate per 100000 persons |
|---|---|---|---|---|---|
| Myocarditis | 938,812 | 21 | 2.24 | 6 | 0.64 |
| Pericarditis | 936,197 | 27 | 2.88 | 18 | 1.92 |

Table 17: Number of reported myocarditis and pericarditis adverse effects and respective reporting rates per 100,000 persons from Israel (source: Barda et al. (2021))

As these are based on persons and not vaccinations the reporting rates should be added and then halved to be somewhat comparable to the German numbers. Then they are still slightly higher but that may be caused by a higher proportion of young adults in the Israelian population compared to the German one. And again at least for the adverse effect myocarditis it can be shown that the rate for the vaccinated group is significantly higher at the 95%-level than for the control group (cf. Table A6 in the appendix). In addition it should be noted that the median age of persons suffering from myocarditis as an adverse effect in the vaccination group was just 25 years.

Further more it is worthwhile noting that Barda et al. (2021) showed that although the relative risk for getting a myocarditis after vaccination with Comirnaty is about 3 times higher than for the control group it is impressively lower than the relative risk increase for obtaining a myocarditis in the course of a SARS-Covid 2-infection which is about 11 times.

According to the security report from the Paul-Ehrlich-Institut (2021-f) there have been nine fatalities in conjunction with a vaccination reported (5 in connection with Comirnaty, 2 with Spikevax [although the myocarditis was not considered as cause], 1 each with Janssen and Vaxzevria, where the causality is still unclear. But even if all cases would be taken into account the overall incidence would be less than 1 case in 10 Millions vaccinations overall, which is close to being negligible compared to the other potential causes of fatalities. However, the situation should be monitored closely as for a lot of cases the final outcome is yet to be determined.

# 7. Summary and Discussion

In this study we analysed the frequencies of adverse events (AE) related to COVID-19 vaccinations in Germany, Austria and Israel using publicly accessible data. Based on statistical analysis we tried to find out, whether the reported frequencies differ with respect to the applied vaccine products. Whenever possible, we discriminated further by gender and age. The reported results relate to AE in general, to severe AE and to several AE of special interest, e.g. thromboembolic side effects.

We categorise the frequencies of AE according to a broadly used classification (cf. e.g. Büchter et al. 2014 or VFA (2021) in Germany).

The reported numbers suggest that adverse events happen at an <u>uncommon</u> frequency, that is 1-10 incidences per 1,000 vaccinations. This is true for both German and Austrian data, although the latter indicates slightly higher incidences.

Severe adverse events can be categorized as <u>rare</u> events, i.e. 1-10 events per 10,000 vaccinations. For these events significant differences among the different vaccines were observed. The reporting rates for mRNA-vaccines were lower compared to those of Janssen and Vaxzevria. The latter ones showed the highest reporting rates.

All above mentioned rates fluctuated somewhat with time but stayed in the same frequency category (uncommon, rare).

There is only poor data available with respect to the gender distribution of the various vaccine products. Taking the simplifying assumption of a homogenous distribution across the vaccines there is a strong indication that the reporting rates for adverse effects are higher for females except for the vaccine Janssen. The assumption of a homogenous distribution might not be valid, as e.g. Austrian data indicate. Nevertheless the relative ranking with respect to reported rates amongst the vaccine was preserved.

Narrowing the focus to incidents with fatal consequences or persisting damages the reporting rates decrease to 0.5 to 2.9 for fatal events and 3.2 to 7.2 for persisting damages per 100,000 vaccinations in Germany. This corresponds mostly to the category "very rare". The findings for fatal events are consistent with results from Austria (there were no data for the category persisting damages available for Austria). Slightly different is the relative ranking for the four vaccine products in both countries. Whilst in Germany Spikevax accumulated the lowest reporting rates for fatal events in Austria it was Janssen. In both countries Vaxzevria showed the highest reporting rates. In Germany Vaxzevria delivered the highest reporting rates for persisting damages as well.

It should be noted that the Paul-Ehrlich-Institut declared that only 78 fatal cases out of 1919 are somewhat likely to be caused by the COVID-19 vaccination (Paul-Ehrlich-Institut (2021-g)). This would indicate that the rate of fatal cases caused by a vaccine itself were about a factor 25 lower.

One category of adverse effects with potentially fatal consequences are thromboembolic events. The provided data from Germany showed that those adverse effects are very rare (except for Vaxzevria which just hit the rare category). The events "venous thrombosis combined with a thrombocytopenia (TTS)" and "ideopathic thrombocytopenia purpura (ITP)" occur even lesser than very rare as their reporting rate is about an order of magnitude lower. The same is true looking at all thromboses associated with a fatal event.

In all considered categories of thrombosis, Vaxzevria showed a significantly worse side effect profile than the other three vaccines products. In particular Spikevax and Comirnaty indicated the lowest risk. Analysing the data from Barda et al. (2021) for Israel it could be concluded that the risks for deep-vein thromboses, other thromboses and thrombocytopenia in the group vaccinated with Comirnaty were even lower than in the control group without vaccination in the same time span.

Another category of high medial interest is the occurrence of myo- and pericarditis. Data from Germany show indeed a significant higher occurrence for these two adverse effects (no differentiation by the Paul-Ehrlich-Institut) assigned to the mRNA-vaccines compared to Vaxzevria. The incidences for Spikevax were significantly higher than for all others, whilst the 95%-confidence interval for Janssen and Comirnaty slightly overlapped. This is in contrast to the findings from Austria were the 95%-confidence intervals for all four vaccines overlapped. The data from Israel however state a clear increased relative risk for myocarditis as an adverse effect for the vaccinated group (solely Comirnaty was applied) compared to the control group. Thankfully the Paul-Ehrlich-Institut provided some more detailed analysis for this hazard. It was shown that the risk is significantly more pronounced for males. Furthermore, primarily younger people are affected. The relative risk ratio for males under 40 years for Spikevax vs. Comirnaty is about 2.5. For females this is similar for the age-group 18-29 with a factor 3. The recommendation of the STIKO (2021-f) to refrain from using Spikevax for persons below 30 years of age seems insofar plausible. However from a relative risk perspective the question should be posed whether this should not be extended to males under 40 years of age as the there seems to be a less risky alternative with Comirnaty, not withstanding that the absolute risk in this age group is only about 40% of that in the age groups 12-29.

It should be noted however that the reporting rates for all age groups lay between 1 and 10 cases per 100,000. i.e. very rare events, except for males between 12-29 with Spikevax as the applied vaccine where the rates increased to 11-12 cases per 100,000 (i.e. rare events).

Looking at the overall situation the four considered vaccines are applied millions if not hundred of millions times worldwide. The observational time span is now more than a year. This actually means probably one of the most thorough field test in medical history. However, there is never an absolute certainty that further adverse effects might be discovered long-term. But from all considered data from three countries the indication seems inevitable that from a statistical point of view vaccinations with all vaccine types provide very low risks of serious undesired consequences. In this respect they can all be regarded very safe, but some are safer than others, but usually not consistently with respect to all potential hazards, e.g. the specific risk profiles for thromboses and myocarditis are different and not homogeneously reflected in data from the three considered countries.

Combining risk considerations with the reported efficacy of the various vaccines (cf. e.g. Harder et al. (2021)) the indication for a preference of the mRNA-vaccines seems obvious not withstanding the positive contribution of the vector-based vaccines in the overall pandemic fight. Whether that conclusion will remain valid as the virus mutates needs to be observed. Fortunately enough the mRNA-platform technology promises rather rapid adaption opportunities of the vaccine as well.

# 8. Further Conclusions

Besides the statistical analyses discussed in the preceding chapter the most surprising and disturbing finding was what could not be analysed due to missing data.

- Concerning the applied vaccinations in Germany the reports of the Paul-Ehrlich-Institut usually do not even provide a stratification of the applied vaccination according to gender and age, let alone relevant other information about specific risk factors of the vaccinated persons, like e.g. a patient history of thromboses[4].

- For the reported incidents of adverse effects information about gender and age stratification is not consistently provided. Furthermore this information of the reported incidents cannot be used for specific risk evaluations.

- The situation for Austria seems by and large similar.

---
4   A request to the Paul-Ehrlich-Institute whether this data could be made available was not specifically answered.

- In Israel apparently a lot of patient specific data was available and used to very carefully model the homogeneity of vaccination and control group as well as the group of infected persons. These data included not only gender and age but also many anamnestic characteristics and even some socio-demographic items. Unfortunately this data was not further exploited to detail the described findings.

Other observations were:

- Not all reported measures are updated in detail in subsequent reports.

- Overall time series for the reported measures are not provided, not even for the most important ones. Hence considerations of trends vs. fluctuations need to be tediously analysed from data gathered from various reports.

- The reports from the Paul-Ehrlich-Institute do not provide exact numbers for certain measures but only percentages rounded to per thousands which leads to quite some rounding uncertainty for absolute numbers.

This leads to the following recommendations for improvement:

- Data from various sources like Paul-Ehrlich and Robert-Koch-Institut and perhaps health administration stakeholders like insurance companies should be pooled in an anonymised and structured way to allow more detailed analyses.

- As part of an open data strategy this data pool should accessible for all parties of genuine interest, e.g. scientists, to allow independent analyses and quality checks. On one hand this would be a measure to instil some more confidence to all people who distrust the government and on the other hand would give concerned people, together with their attending physicians, the opportunity to choose from the set of various and generally very safe vaccines the one that seems most appropriate for their personal risk situation.

- The reports from the Paul-Ehrlich-Institut should be enriched by using the more holistic data when available (cf. above). They should be standardised in so far that key characteristics are reported in the same format in each report and time series should be provided to easily monitor developments and recognise effects of e.g. altered vaccination strategies.

# Appendix – Tables

|  | Number of applied vaccination doses | Total reported events | Respective reporting rate per 1000 vaccinations | Lower bound of 95%-confidence interval | Upper bound of 95%-confidence interval |
|---|---:|---:|---:|---:|---:|
| Comirnaty | 11126627 | 16262 | 1.46 | 1.439 | 1.484 |
| Spikevax | 1277515 | 3704 | 2.9 | 2.807 | 2.994 |
| Vaxzevria (AZ) | 1578469 | 18818 | 11.92 | 11.754 | 12.091 |
| Covid19-Vaccine Janssen | 344801 | 1097 | 3.18 | 2.997 | 3.375 |
| Total | 14327412 | 39881 | 2.78 | | |

*Table A1: Number of reported adverse effects for* Austria including 95%-confidence intervals for the reporting rate for adverse effects per 1,000 vaccinations. *Data from 27.12.2020 until 3.12.2021 (BASG 2021).*

|  | Gender | Number of applied vaccination doses | Total number of applied vaccination doses | Percentage |
|---|---|---:|---:|---:|
| Comirnaty | Female | 5782499 | 11123804 | 52.0 |
|  | Male | 5341305 | 11123804 | 48.0 |
| Spikevax | Female | 612664 | 1277108 | 48.0 |
|  | Male | 664444 | 1277108 | 52.0 |
| Vaxzevria (AZ) | Female | 853721 | 1578008 | 54.1 |
|  | Male | 724287 | 1578008 | 45.9 |
| Covid19-Vaccine Janssen | Female | 133876 | 344548 | 38.9 |
|  | Male | 210672 | 344548 | 61.1 |

*Table A2: Number of applied vaccinations per vaccine and gender in Austria. Data from 27.12.2020 until 3.12.2021 (BMSGPK 2021).*

|  | Number of applied vaccines | Cumulated number of deaths | Corresponding reporting rate | Lower bound of 95%-confidence interval | Upper bound of 95%-confidence interval |
|---|---|---|---|---|---|
| Comirnaty | 11126627 | 151 | 1.36 | 1.149 | 1.592 |
| Spikevax | 1277515 | 18 | 1.41 | 0.835 | 2.227 |
| Vaxzevria (AZ) | 1578469 | 30 | 1.9 | 1.282 | 2.713 |
| Covid19-Vaccine Janssen | 344801 | 3 | 0.87 | 0.179 | 2.543 |
| Totals | 14327412 | 202 | 1.41 | | |

Table A3: Number of reported deaths and their respective reporting rates per 100,000 applied vaccinations, including the 95%-confidence intervals for the reporting rates, for Austria. Data from 27.12.2020 until 3.12.2021 (BASG 2021).

| Number of reported adverse events of thrombosis with or without thrombocytopenia and respective reporting rates from 27.12.2020 until 1.8.2021 | Total number of applied doses | Number of thromboses without thrombocytopenia | Respective reporting rate per 100000 vaccinations | Number of thromboses without thrombocytopenia and with fatal consequences | Respective reporting rate per 100000 vaccinations | Number of thromboses with thrombocytopenia | Respective reporting rate per 100000 vaccinations | Number of thromboses with thrombocytopenia and with fatal consequences | Respective reporting rate per 100000 vaccinations | Total number of thromboses | Respective reporting rate per 100000 vaccinations | Total number of thromboses with fatal consequences | Respective reporting rate per 100000 vaccinations | Number of thrombocytopenia without thrombosis | Respective reporting rate per 100000 vaccinations | Number of thrombocytopenia without thrombosis and with fatal consequences | Respective reporting rate per 100000 vaccinations |
|---|---|---|---|---|---|---|---|---|---|---|---|---|---|---|---|---|---|
| Comirnaty | 68,962,481 | 1,758 | 2.55 | 181 | 0.26 | 8 | 0.01 | 1 | 0.00 | 1,766 | 2.56 | 182 | 0.26 | 172 | 0.25 | 8 | 0.01 |
| Spikevax | 8,506,260 | 185 | 2.17 | 10 | 0.12 | 2 | 0.02 | 0 | 0.00 | 187 | 2.20 | 10 | 0.12 | 16 | 0.19 | 0 | 0.00 |
| Vaxzevria (AZ) | 12,491,937 | 1,264 | 10.12 | 66 | 0.53 | 174 | 1.39 | 27 | 0.22 | 1,438 | 11.51 | 93 | 0.74 | 206 | 1.65 | 6 | 0.05 |
| Janssen | 2,416,109 | 69 | 2.86 | 4 | 0.17 | 12 | 0.50 | 4 | 0.17 | 81 | 3.35 | 8 | 0.33 | 15 | 0.62 | 1 | 0.04 |
| Totals | 92,376,787 | 3,207 | 3.47 | 247 | 0.27 | 184 | 0.20 | 28 | 0.03 | 3,391 | 3.67 | 275 | 0.30 | 394 | 0.43 | 14 | 0.02 |

Table A4: various figures on thromboses calculated using data from PEI 2021-d

| 95%-confidence limits for reporting rates for ITP and fatal events thereof (data until 1.8.2021) | lower limit all ITP per 100000 vaccinations | upper limit all ITP per 100000 vaccinations | lower limit all ITP with fatal ending per 100000 vaccinations | upper limit all ITP with fatal ending per 100000 vaccinations |
|---|---|---|---|---|
| Comirnaty | 0.214 | 0.290 | 0.005 | 0.023 |
| Spikevax | 0.108 | 0.305 | 0.000 | 0.043 |
| Vaxzevria (AZ) | 1.432 | 1.890 | 0.018 | 0.105 |
| Janssen | 0.347 | 1.024 | 0.001 | 0.231 |

Table A5: 95%-confidence limits for reporting rates for thrombocytopenia without thrombosis (ITP) and fatal events thereof (data until 1.8.2021)

|  | Number of applied vaccines | Number of myo-/pericarditis adverse effects | Corresponding reporting rate | Lower bound of 95%-confidence interval | Upper bound of 95%-confidence interval |
|---|---|---|---|---|---|
| Comirnaty | 11126627 | 122 | 1.10 | 0.911 | 1.309 |
| Spikevax | 1277515 | 18 | 1.41 | 0.835 | 2.227 |
| Vaxzevria (AZ) | 1578469 | 16 | 1.01 | 0.579 | 1.646 |
| Covid19-Vaccine Janssen | 344801 | 7 | 2.03 | 0.816 | 4.183 |
| Totals | 14327412 | 163 | 1.14 | | |

Table A6: Number of reported cases of myo-/pericarditis and their respective reporting rates per 100,000 applied vaccinations, including the 95%-confidence intervals for the reporting rates, for Austria. Data from 27.12.2020 until 3.12.2021 (BASG 2021).

| 95%-confidence interval limits of reporting rates for myo-/pericarditis per 100000 vaccinations (data 27.12.2020 until 30.9.2021) | lower limit per 100000 vaccinations | upper limit per 100000 vaccinations |
|---|---|---|
| Comirnaty | 1.058 | 1.204 |
| Spikevax | 2.159 | 2.795 |
| Vaxzevria (AZ) | 0.368 | 0.617 |
| Janssen | 0.558 | 1.233 |
| Totals | 1.101 | 1.230 |

Table A7: 95%-confidence interval limits of reporting rates for myo-/pericarditis per 100,000 vaccinations (data 27.12.2020 until 30.9.2021)

| | males | | females | |
|---|---|---|---|---|
| 95%-confidence interval limits of reporting rates for myo-/pericarditis per 100000 vaccinations per sex (data 27.12.2020 until 30.9.2021) | lower limit per 100000 vaccinations | upper limit per 100000 vaccinations | lower limit per 100000 vaccinations | upper limit per 100000 vaccinations |
| Comirnaty | 1,499 | 1,752 | 0,599 | 0,757 |
| Spikevax | 3,413 | 4,581 | 0,656 | 1,401 |

Table A8: 95%-confidence interval limits of reporting rates for myo-/pericarditis per 100,000 vaccinations differentiated by gender (data 27.12.2020 until 30. 9. 2021)

| 95%-confidence interval limits of reporting rates for myo-/pericarditis per 100000 persons | lower limit per 100000 persons vaccinated | upper limit per 100000 persons vaccinated | lower limit per 100000 persons in control group | upper limit per 100000 persons in control group |
|---|---|---|---|---|
| Myocarditis | 1.385 | 3.419 | 0.249 | 1.476 |
| Pericarditis | 1.901 | 4.196 | 1.206 | 3.215 |

Table A9: 95%-confidence interval limits of reporting rates for myo-/pericarditis per 100,000 vaccinations differentiated by vaccination vs. control group (data source Barda et al. -2021)